\documentclass[letterpaper,10pt]{article}
\usepackage{osameet2}

\usepackage{amsmath,amssymb}
\usepackage[T1]{fontenc}
\usepackage{graphicx}
\usepackage{subfigure}
\usepackage{wrapfig}
\usepackage{amsfonts}
\usepackage{footmisc}
\usepackage{scrextend}
\usepackage{multirow}
\usepackage{cleveref}
\usepackage{indentfirst}
\usepackage{braket}
\usepackage{float}
\usepackage{amsmath}
\usepackage{epstopdf}
\usepackage{esint}
\usepackage{color}
\usepackage[T1]{fontenc}
\usepackage{subfigure}
\usepackage{caption}

\DeclareMathAlphabet      {\mathbf}{OT1}{cmr}{bx}{n}

\usepackage[pdftex,colorlinks=true,bookmarks=false,citecolor=blue,urlcolor=blue]{hyperref} 

\begin{document}

\title{Large-Alphabet Encoding Schemes for Floodlight Quantum Key Distribution}

\author{Quntao Zhuang$^{1,2}$ \ Zheshen Zhang$^1$ \ Jeffrey H. Shapiro$^1$}
\address{$^1$Research Laboratory of Electronics, Massachusetts Institute of Technology, Cambridge, Massachusetts 02139, USA\\
$^2$Department of Physics, Massachusetts Institute of Technology, Cambridge, Massachusetts 02139, USA}
\email{email: quntao@mit.edu}

\begin{abstract}
Floodlight quantum key distribution (FL-QKD) uses binary phase-shift keying (BPSK) of multiple optical modes to achieve Gbps secret-key rates (SKRs) at metropolitan-area distances.  We show that FL-QKD's SKR can be doubled by using $32$-ary PSK.
\end{abstract}

\vspace*{.32in}

Quantum key distribution~\cite{Bennett20147} (QKD) allows remote parties (Alice and Bob) to create a shared random bit string with unconditional security.  Later, they can employ their shared string for one-time-pad (OTP) encryption of messages they wish to keep entirely private from any eavesdropper (Eve).  Unfortunately, current QKD systems' secret-key rates (SKRs) fall far short of what is needed to make high-speed (Gbps) transmission with OTP encryption ready for widespread deployment.  Floodlight QKD (FL-QKD)~\cite{Quntao_2015,Zhang_2016} is a new protocol that uses binary phase-shift keying (BPSK) of multiple optical modes and homodyne detection to achieve security against the optimum frequency-domain collective attack.  It is predicted to permit Gbps SKRs at metropolitan-area distances in a single-wavelength implementation without the need to develop any new technology.  In this paper we extend FL-QKD's security analysis to $K$-ary phase-shift keying (KPSK), and show that the increased alphabet size affords SKR increases by up to a factor of two.  Thus, over a 50-km-long fiber, going from BPSK to 32-ary PSK increases FL-QKD's SKR from 2.0\,Gbps to 4.5\,Gbps. 

\begin{wrapfigure}[17]{l}{0.5\textwidth}
\includegraphics[width=0.5\textwidth]{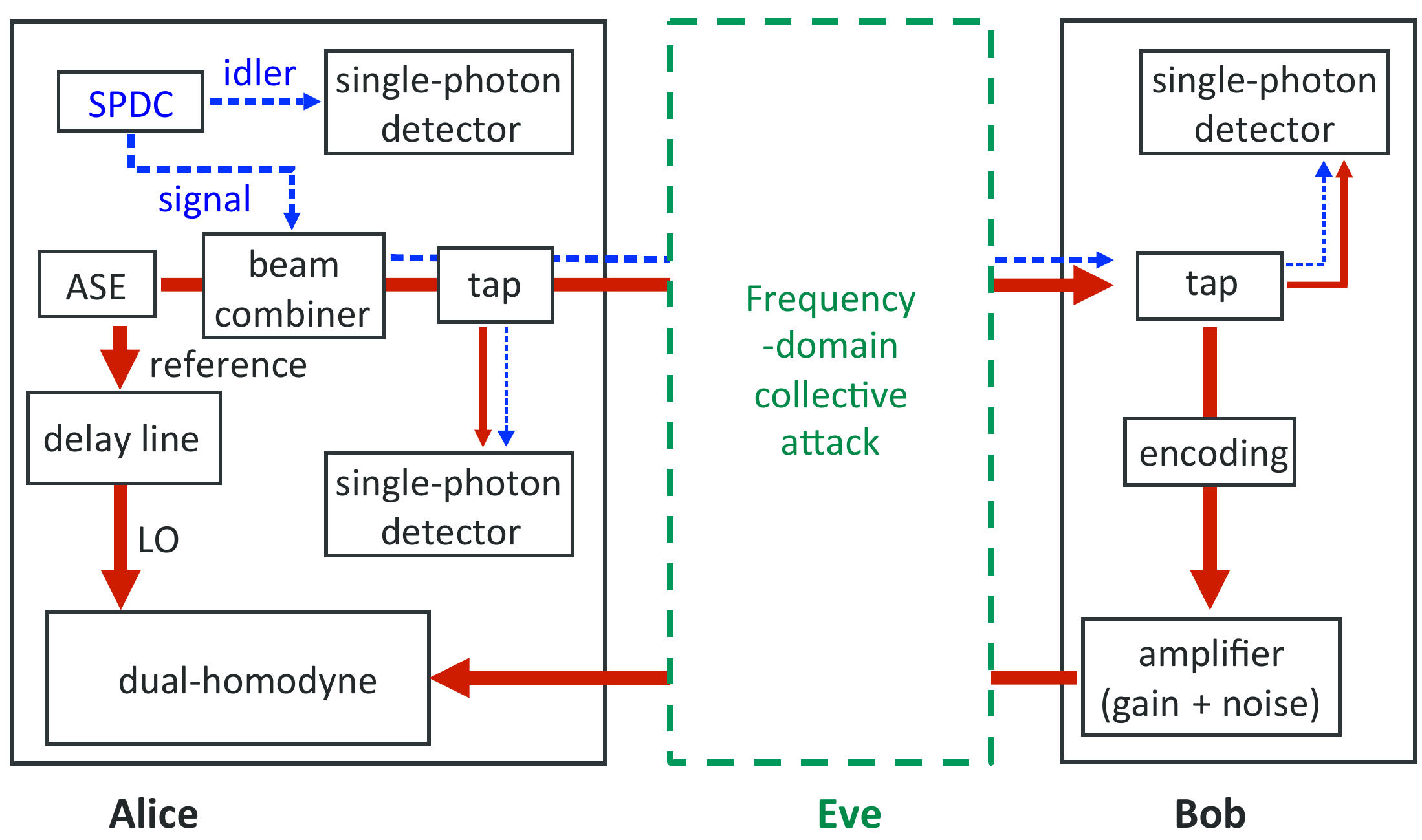}
\caption[caption]{Quantum channel setup for FL-QKD.  ASE:  amplified spontaneous emission source.  SPDC:  spontaneous parametric downconverter. LO:  local oscillator.}
\label{scheme_FLQKD}
\end{wrapfigure}

In KPSK FL-QKD~(schematic shown in Fig.~\ref{scheme_FLQKD}), Alice splits the $W$-Hz-bandwidth, flat-top spectrum, high-brightness output from an amplified spontaneous emission~(ASE) source into a low-brightness signal and a high-brightness reference.  To enable channel monitoring, Alice combines her low-brightness ASE in an $n\gg 1$ ASE-to-SPDC-ratio with the signal output from a spontaneous parametric downconverter (SPDC) of the same $W$-Hz-bandwidth flat-top spectrum.  Alice uses a single-photon detector to monitor her SPDC's idler and another single-photon detector to monitor a $\kappa_A \ll 1$ fraction of her combined ASE-SPDC light, while sending the remainder of that light---whose brightness is $N_S \ll 1$ photons/mode---to Bob.  Alice retains her bright reference beam in an optical-fiber delay line---using amplifiers as needed---for use as her dual-homodyne receiver's $N_{\rm LO} \gg 1$ photons/mode brightness local oscillator (LO).   

In the absence of Eve, the fiber link from Alice to Bob is a pure-loss channel with transmissivity $\kappa_S\ll1$. Eve's presence, however, allows her control that channel, hence  Alice and Bob must perform channel monitoring to bound Eve's information gain. So, prior to his KPSK encoding operation, Bob routes a small fraction $\kappa_B\ll 1$ of the light he receives to a single-photon detector. The outputs from Alice and Bob's single-photon detectors enable them to determine the single rates $S_I$ for Alice's idler and $S_A\, \left(S_B\right)$ for Alice's~(Bob's) tap, as well as $C_{IA}\, \left(C_{IB}\right)$ and $\widetilde{C}_{IA}\, \left(\widetilde{C}_{IB}\right)$, the time-aligned and time-shifted coincidence rates between Alice's idler and Alice's~(Bob's) tap.  They use their measurements to:  (1) verify that Bob receives the photon flux he would get were Eve absent; and (2) determine Eve's intrusion parameter
$
f_E = 1-[(C_{IB}-\widetilde{C}_{IB})/S_B]/[(C_{IA}-\widetilde{C}_{IA})/S_A],
$
which quantifies the integrity of the Alice-to-Bob channel and allows them to place an upper bound, $\chi_{EB}^{\rm UB}$, on Eve's Holevo-information rate for her optimized frequency-domain collective attack, which she can realize in the form of an SPDC light-injection attack~\cite{Quntao_2015}.

Bob's KPSK modulation works as follows.  In each $T$-s-duration symbol interval (symbol rate $R = 1/T$), Bob applies a $2\pi k/K$ phase shift to the light remaining after his monitor tap, where $k$ is equally likely to be any integer between 0 and $K-1$ and the $k$ values for different symbol intervals are statistically independent.   He then amplifies his modulated light with a gain $G_B  \gg 1$ amplifier whose output ASE has brightness $N_B = G_B -1$, and sends the amplified and modulated light back to Alice through what, in Eve's absence, is a $\kappa_S$-transmissivity fiber.  The amplifier's gain will overcome the return-path loss insofar as Alice is concerned, while its output ASE will mask Bob's modulation from Eve.  To decode Bob's symbols, Alice uses dual-homodyne reception, i.e., she 50--50 beam splits both the light returned from Bob and her LO, and then makes homodyne measurements of the $I$ (0 phase shift) and $Q$ ($\pi/2$ phase shift) in-phase and quadrature components of the returned light as in classical KPSK fiber-optic communication.  
\begin{wrapfigure}[19]{L}{0.6\textwidth}
\subfigure{
\includegraphics[width=0.27\textwidth]{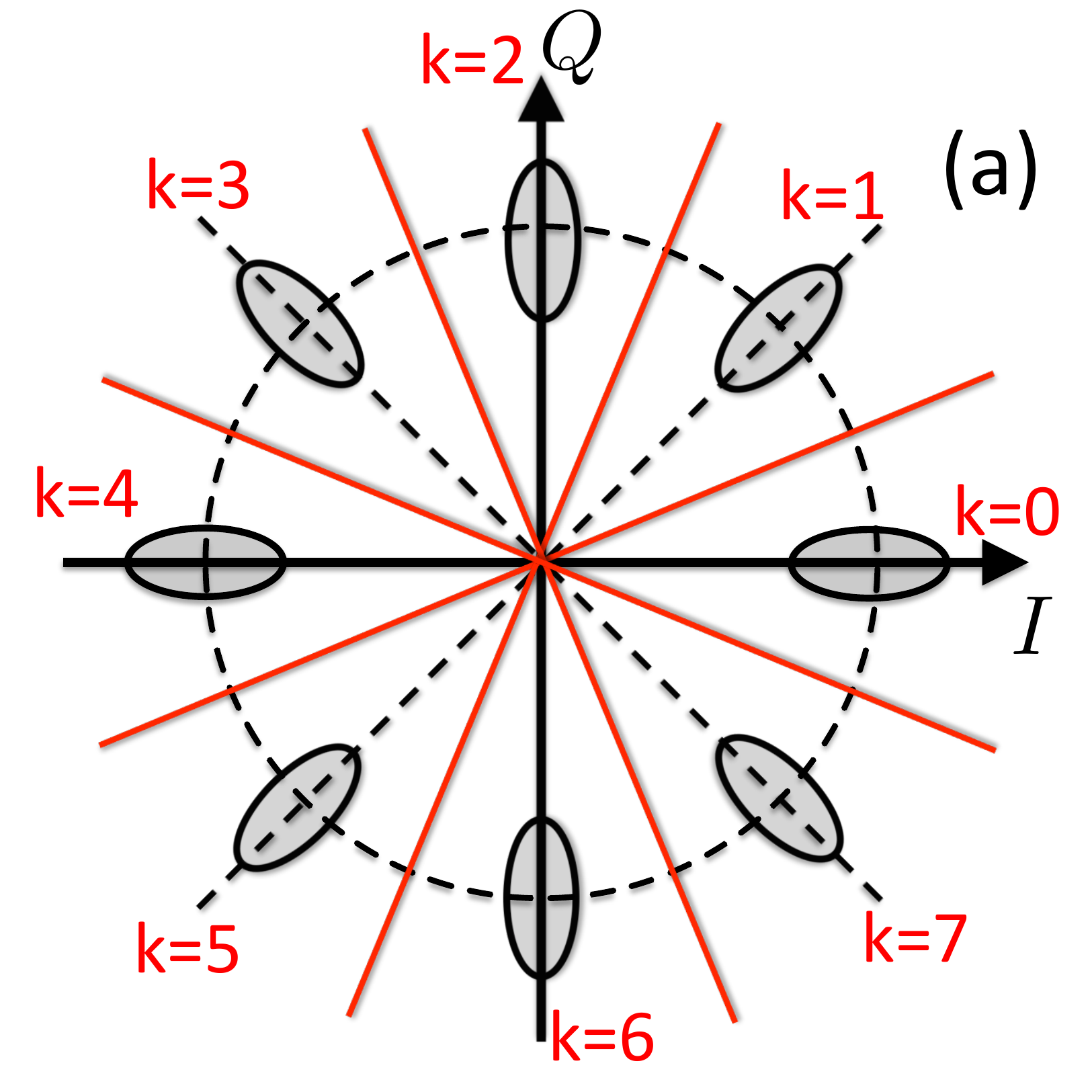}
\label{scheme_modulation}
}
\subfigure{
\includegraphics[width=0.27\textwidth]{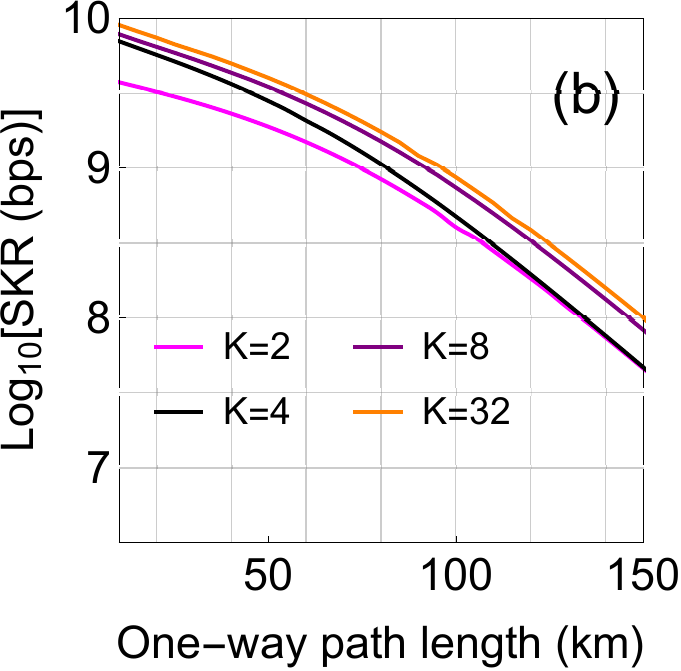}
\label{rate}
}
\caption[caption]{(a) 8-ary PSK example.  The gray-shaded ellipses enclose one-standard-deviation regions for Alice's receiver about the $I+iQ$ values of Bob's symbols, and the red lines mark the boundaries of her minimum error-probability decision regions.  (b) SKR $\Delta I_{AB}^{\rm LB}$ vs. one-way path length.  At 50\,km , $\Delta I_{AB}^{\rm LB}\approx 2.0$ and 4.5\,Gbps for $K=2$ and 32.
}
\end{wrapfigure}

When $M = TW$, the number of optical modes per symbol, is high ($M\gg 1$),
the joint statistics of $I$ and $Q$ conditioned on knowledge of $k$ can be well approximated by a Gaussian distribution whose symmetric behavior for $0\le k \le K-1$ is shown, schematically, in Fig.~\ref{scheme_modulation}.  This symmetry, plus all $k$ values being equally likely, makes Alice's minimum error-probability decision rule choosing her decoded symbol to be the one whose signal location in the $I$-$Q$ plane is closest to her measured $(I,Q)$ value.  See~Fig.~\ref{scheme_modulation} for the resulting decision regions.  

Once Alice has decoded Bob's string of transmitted symbols the two of them use a tamper-proof classical channel (not shown in Fig.~\ref{scheme_FLQKD}) to perform reconciliation (error correction) and privacy amplification.  During reconciliation, Alice and Bob obtain values for the conditional probabilities $\Pr(\,\tilde{k}\mid k\,)$, i.e., the probabilities that Alice decoded $\tilde{k}$ given Bob sent $k$, from which they calculate their Shannon-information rate $I_{AB}$ via
$
I_{AB}=R\{\sum_{k=0}^{K-1}\sum_{\tilde{k}=0}^{K-1}[\Pr(\,\tilde{k}\mid k\,)/K]\log_2[K\Pr(\,\tilde{k}\mid k\,)/\sum_{k'=0}^{K-1}\Pr(\,\tilde{k}\mid k'\,)]\}.
$
Then, using their upper bound on Eve's Holevo-information rate, they know that their achievable SKR is bounded from below by
$
\Delta I_{AB}^{\rm LB} = \beta I_{AB} - \chi_{EB}^{\rm UB},
$
where $\beta$ is the efficiency of their reconciliation algorithm, and, because of FL-QKD's extraordinarily high SKR, finite-key effects have been neglected.  

To explore the SKR behavior of KPSK FL-QKD we performed numerical maximization of $\Delta I_{AB}^{\rm LB}$ over Alice's source brightness, $N_S$, for one-way path lengths up to 150\,km using parameter values similar to those employed in Ref.~\cite{Quntao_2015}: $W=2$\,THz source bandwidth; $n=99$ ASE-to-SPDC ratio; $\kappa_A = \kappa_B = 0.01$ monitor taps;  $0.2$\,dB/km fiber loss; $R=10$\,Gbaud symbol rate; $G_B = 10^6$ amplifier gain; $N_{\rm LO} = 10^4$ LO brightness; 0.9 homodyne-detection efficiency; and $\beta=0.94$ reconciliation efficiency.  The maximum SKRs we obtained for $K=2$, 4, 8, and 32 are shown in Fig.~\ref{rate}.  We see that going from BPSK to 32-ary PSK approximately doubles the achievable SKR over all the distances shown, with BPSK providing 2.0\,Gbps SKR and 32-ary PSK giving 4.5\,Gbps SKR at 50\,km.   

It is interesting to note how FL-QKD's KPSK performance differs from that seen in fiber-optic communication using high-order signal constellations and coherent detection~\cite{Li_2009}.  In fiber-optic communication, high-order signal constellations can enormously improve spectral efficiency (bits/sec-Hz = bits/mode), and such systems are now moving beyond KPSK to quadrature amplitude modulation (QAM).  Our work shows that FL-QKD benefits from the increased spectral efficiency of KPSK, but we have found that there is no value to conventional (square-lattice) QAM, because that format's amplitude modulation gives away too much information to Eve.

\end{document}